\documentclass[proceedings, preprint]{rmaa}

\usepackage{paralist}

\usepackage{psfrag,color}

\SetYear{2015}
\SetConfTitle{Fourth Workshop on Autonomous Observatories}

\title{Autonomous Dome for Robotic Telescope} 

\author{
  Akash Kumar,\altaffilmark{1} 
  Anand Sengupta,\altaffilmark{1}
  and Shashikiran Ganesh\altaffilmark{2}}

\altaffiltext{1}{Indian Institute of Technology, Village Palaj, Shimkhera, Gandhinagar, Gujarat, India-382355 (asengupta@iitgn.ac.in).}

\altaffiltext{2}{Physical Research Laboratory, Navrangpura, Ahmedabad, Gujarat, India  (shashi@prl.res.in)}

\shortauthor{Kumar, Sengupta, \& Ganesh}
\shorttitle{Fourth Workshop on Autonomous Observatories}

\listofauthors{Akash Kumar, Anand Sengupta, \& Shashikiran Ganesh}
\indexauthor{Kumar, A.}
\indexauthor{Sengupta, A.}
\indexauthor{Ganesh, S.}

\abstract{
Physical Research Laboratory operates a 50cm robotic observatory at Mount Abu.  This Automated Telescope for Variability Studies (ATVS) makes use of Remote Telescope System 2 (RTS2) for autonomous operations.  The observatory uses a 3.5m dome from Sirius Observatories.  We have developed electronics using Arduino electronic circuit boards with home grown logic and software to control the dome operations.  
We are in the process of completing the drivers to link our Arduino based dome controller with RTS2.  
This document is a short description of the various phases of the development and their integration to achieve the required objective.}

\addkeyword{Arduino}
\addkeyword{Dome Automation}
\addkeyword{Robotic observatory}
\addkeyword{Autonomous observatory}

\begin{document}
\maketitle

\section{Introduction}
\label{sec:intro}

Physical Research Laboratory operates a robotic 50cm telescope, Autonomous Telescope for Variability Studies(ATVS), at its observatory at Mount Abu (latitude : $24^\circ 39^m 9^s$ North, longitude: $72^\circ 46^m 47^s$ East, altitude : $1680 m$), India.   The operating conditions are quite tough in terms of range in humidity, temperature and wind.   The telescope is protected from the environment by a fibre glass dome manufactured by Sirius Observatories.  \\

\section{Dome controller for autonomous operations}

The telescope operates in the robotic mode\citep{Ganesh2013}  and efforts are on way to fully automate it using Remote Telescope System (RTS2, \url{http://rts2.org/}).   The observatory has over 200 clear nights in a year but needs to be completely closed and sealed during the Indian monsoon season (generally mid June to September) every year.  Apart from this there are occasions when the skies become cloudy and/or wind conditions also increase drastically.  For autonomous operations to succeed, therefore, we need to have the dome controller be very reliable with independent inputs/monitoring of the weather conditions so that the shutters could be closed automatically in case of bad weather.   \\

Humidity being a major cause of repeated failures of electronic boards, it has not been found feasible to use commercially available solutions for controlling the dome.  Hence we decided to go for a locally engineered solution using cheap, general purpose, electronic boards available locally.  The objective of this  work  was to make an autonomous dome controller for a robotic telescope using easily replaceable electronics. 
The controller is in charge of the dome functions like closing and opening of dome's shutter, clockwise and counter-clockwise motion of the dome to track the telescope.  It also keeps the record of the position of the dome and can be controlled by a Windows or Linux based computer using appropriate drivers.  \\


We built the dome controller using the ubiquitous Arduino boards.  An Arduino is an
open source physical computing platform based on a simple micro-controller
board. It consists of a physical programmable circuit board.  It can be programmed via the USB port of the computer using the Arduino IDE(integrated Development Environment) built with the Processing platform. The Arduino 
IDE includes support for various electronic components such as encoders and other sensors, relay boards etc. We
used one Arduino board to control the dome's shutter movement (opening
and closing) and another one to control the dome rotation (clockwise or 
counter-clockwise).  Both the boards communicate with each other using 
RF transceivers.  This is necessitated because the dome rotation controller is connected to the PC and the shutter controller is connected via the dome controller.  The shutter controller, powered by a battery,  is mounted in a box on the dome and rotates with the dome.  Thus wireless communication is a must.    For monitoring the orientation of the dome we use an incremental rotary encoder which  
converts the angular motion of the dome into a series of digital pulses which encode the movement of the dome.  
To control the motion of motors of the shutter and dome we have used semiconductor relay boards with physical limit switches .  These semiconductor relay boards are operated using digital pulses from the Arduino microcontroller board. 
\\
 \section{Drive Development Phases}
\label{sec:errors}
\begin{figure}[!t]
\includegraphics[width=\columnwidth]{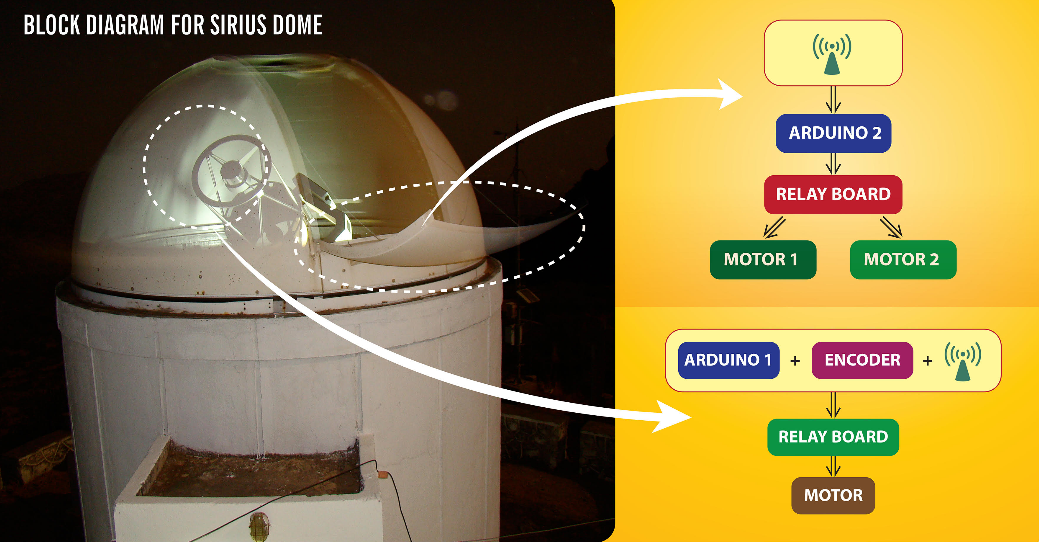} 
  \caption{View of Sirius Dome housing the PRL 50cm telescope at Mount Abu (left panel).  Block diagram of dome and shutter control logic (right panel).}
  \label{fig:simple}
\end{figure}

 The work was divided in different phases and work of all the phases was accomplished one by one and finally integrated to make the drive. The different phases were\\ (i) Making a  programmable controller of the dome's shutter \\(ii) Making a programmable controller of the dome's rotation \\(iii) Conversion of  the dome's motion into digital code \\(iv) Setting up a wireless connection between the dome's rotation controller and the dome's shutter\\
(v) Making it compatible with RTS2.\\ 

 In the first phase, a mechanical controller circuit for the dome's shutter was made using limit switches and was successfully tested. After that a programmable control circuit for the dome's shutter was made using an arduino board and semiconductor relays to control the motors and was coded  for opening, intermediate stop and closing operations of the shutter. Work on the position encoder was also started to record the dome's position.\\
 
 Simultaneously, work on the rotation part of the dome was started and a programmable control circuit was made using another Arduino board and was coded for clockwise, counter-clockwise motion and for sensing home position.  
 Then the position encoder was integrated with this controller and the code was modified accordingly. Work on the wireless communication between rotation and shutter controllers was started using radio frequency transceivers and a communication protocol was established and encoded for transferring the command to the controller boards.  \\
 
 The electronic circuits have been tested at the observatory and software coding has been developed for synchronizing the dome orientation with the azimuth being pointed by the telescope.  We plan to integrate the shutter control board with independent cloud, rain and wind sensors to allow quick closure of the shutters independent of any computer control / manual intervention.  \\
 
 {\em Acknowledgement}  {\bf The first author would like to acknowledge support received from the organizers towards local hospitality which enabled his participation in the workshop.  His travel to Malaga was funded by IIT Gandhinagar.  Work at PRL is funded by the Dept of Space, Govt. of India.  We thank colleagues at PRL for their support of this work.}

\end{document}